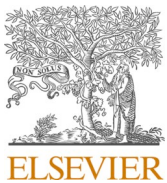
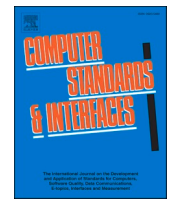

# A microservice architecture for real-time IoT data processing: A reusable Web of things approach for smart ports

Guadalupe Ortiz [a], Juan Boubeta-Puig [a,*], Javier Criado [b], David Corral-Plaza [a], Alfonso Garcia-de-Prado [a], Inmaculada Medina-Bulo [a], Luis Iribarne [b]

[a] *UCASE Software Engineering Research Group, University of Cadiz, Avda. de la Universidad de Cádiz 10, 11519 Puerto Real, Cádiz, Spain*
[b] *Applied Computing Group, University of Almeria, Carretera Sacramento s/n, 04120 La Cañada de San Urbano, Almería, Spain*

ABSTRACT

Major advances in telecommunications and the Internet of Things have given rise to numerous smart city scenarios in which smart services are provided. What was once a dream for the future has now become reality. However, the need to provide these smart services quickly, efficiently, in an interoperable manner and in real time is a cutting-edge technological challenge. Although some software architectures offer solutions in this area, these are often limited in terms of reusability and maintenance by independent modules —involving the need for system downtime when maintaining or evolving, as well as by a lack of standards in terms of the interoperability of their interface. In this paper, we propose a fully reusable microservice architecture, standardized through the use of the Web of things paradigm, and with high efficiency in real-time data processing, supported by complex event processing techniques. To illustrate the proposal, we present a fully reusable implementation of the microservices necessary for the deployment of the architecture in the field of air quality monitoring and alerting in smart ports. The performance evaluation of this architecture shows excellent results.

## 1. Introduction

The Internet of things (IoT) [1] has grown enormously in significance over recent years, mainly due to the great advance in technologies and the capacity to develop smaller and less expensive electronic components. This, together with more affordable communications in general, and improved wireless communications in particular, has encouraged the development of smart cities [2].

Smart cities collect data from multiple sources: sensors, government and business databases (e.g., transportation companies), social networks, data provided by citizens through apps, etc. Thanks to the processing of these data and the taking of appropriate actions, the proper functioning of city's services can be improved and the quality of citizens' lives can thus be enhanced. The scope of application is immense, from basic services, such as garden care [3], or waste collection [4], to personalized and contextualized alerts [5].

Despite the fact that an increasing number of cities are offering new services within the scope of smart cities, technological challenges are opening up to achieve a better and faster service [6]. Current initiatives have failed to establish an interoperability layer for communication and data processing since they are not built according to current standards related to this domain [7,8]. As an example, an abstraction layer would definitely facilitate both the understanding among different parties and the addition of new data sources and data sinks on demand. In many cases, these data sources will be heterogeneous and transmitted through various protocols, which will entail the data processing system having to homogenize the data before processing them and having to support various input protocols. It is difficult, however, to develop a scalable and maintainable system if we wish to add new data sources in the future without knowing what formats and protocols will be used, and without having to stop the system. The same applies to notifications or alerts issued by the system once situations of interest are detected after data processing: each third-party interested in such notifications may require different output formats and communications protocols, and recipients should be able to join the system at any time without the service needing to be stopped. Moreover, the current demand for these services is in real time, and therefore the system must be able to process such data in streaming to be able to notify the detected situation of interest in real time.

In addition to the problem of data processing, there is a need to standardize the services exposed by IoT devices and other elements of the software architectures for IoT. In this sense, the Web of things (WoT) paradigm [7] homogenizes the access to these services by stablishing a standardized mechanism to describe and implement the Application Programming Interfaces (APIs) of such services. The construction of structured APIs based on web technologies creates an abstract layer that enables and supports the interoperability of the system. Thus, it enables the implementation of a variety of components (i.e., things) that offer different alternatives for similar services that can be discovered and used dynamically, depending on the context. The discovery of things requires they are formally described to expose their interaction capacities. WoT Thing Description (TD) [7] is the standard to describe things in terms of three interaction affordances (properties, actions and events), and allows us to build a repository of alternative component definitions. Therefore, these definitions can be inspected to find the most suitable components, for building or adapting a software architecture at runtime [9].

Some of the authors of this paper have made a number of proposals that, although not specific to smart cities, enable real-time data processing and notification in various IoT domains, including smart cities [10–13]. To this end, we have proposed a Complex Event Processing (CEP) layer [14] on the architecture application layer for the processing, correlation and analysis of big data in real time in order to automatically detect situations of interest for a particular domain. This requires the definition and implementation of event patterns that indicate the necessary conditions to detect such situations. One of the great benefits of this technology is its ability to react quickly to critical situations and take actions that automatically respond to them. Unlike other approaches for data analysis such as knowledge fusion [15] (offline processing), the CEP technology requires no prior storage to be able to process data (online processing). Furthermore, some of the authors of this paper have integrated CEP with Event-Driven Service-Oriented Architecture (ED-SOA or SOA 2.0) [10,11], a software architecture characterized by the fact that communications between users, applications and services are carried out by means of events in a totally decoupled way. Such an integration facilitates the decision-making process by automatically sending contextual alerts on detected situations of interest to stakeholders, such as monitoring consoles, intelligent mobile devices and cyber-physical devices.

However, these architectures have two limitations. On the one hand, the one previously described for smart city software architectures in general: despite their being scalable and maintainable, when it is necessary to add new heterogeneous data sources that come with different data formats than those supported by the system or in different protocols than those included in the system, the service must be stopped to add the new elements [12]. On the other hand, although some proposed SOAs 2.0 can be reused in application domains other than that in which a particular architecture has been tested, system-independent modules, such as those for data acquisition, data processing, and notification of situations of interest, cannot be reused without previous software engineering work [16].

In [17], some of the authors of this paper already made progress in the direction of migrating from a full monolithic SOA 2.0 architecture to a microservice one. Consequently, we obtained benefits in the question of architecture maintenance and evolution, but the reuse of the different modules was still limited due to their being coupled to the communication protocols used and the data formats processed by the CEP, as well as the use of an Enterprise Service Bus (ESB). Additionally, we did not provide the TD for the microservices.

In this paper, we seek to go a step further with the aim of developing several fully reusable software components to process, correlate and analyze smart city data to detect and notify situations of interest in real time through the implementation of a microservice-based architecture composed of independent components, as indicated above. More specifically, this architecture will benefit from the use of both a CEP engine in the component responsible for data processing to detect situations of interest, and WoT for an easier understanding and reuse of the description of the proposed components. Therefore, the main contributions of the paper are as follows: Firstly, an efficient microservice architecture for real-time processing of vast and heterogeneous amounts of data from different domain applications in the scope of smart IoT and WoT. Secondly, a total implementation of the microservice architecture that is reusable for any scenario of air quality monitoring and alerting in the context of smart ports. Finally, the configuration, deployment and thing description (TD), together with the evaluation of the case study, proving its usefulness for air quality monitoring and alerting in smart ports. Remarkably, although the proposed microservice architecture is adequate for IoT scenarios considering the low computational power, limited memory and bandwidth imposed by this paradigm, the architecture can be also applied in other non-IoT scenarios which provide more computational power and resources.

The rest of the paper is organized as follows. Section 2 provides the background on the technologies used in the paper. Section 3 describes the proposed architecture in depth, while Section 4 goes into the details of the particular microservices implemented to build the generic architecture in the smart port domain. Section 5 describes the case study used to evaluate the proposed architecture and the results obtained in the evaluation. Related work is then discussed in Section 6. Finally, discussion and conclusions are presented in Section 7 and Section 8, respectively.

## 2. Background

This section describes the background on microservices in the WoT, CEP and messaging in smart scenarios.

### 2.1. Microservices in the Web of things

The microservice architecture pattern has emerged as an alternative to monolithic architectures and applications [18], which are difficult to maintain and evolve due to the high coupling among their components. A microservice architecture is based on the concept of building an application as a set of small interconnected services, which communicate through light protocols [19].

In a microservice architecture, each service is expected to implement an identified function of the server application and should have access to its own database [20]. Furthermore, the communications between services should be based on Hypertext Transfer Protocol/ REpresentational State Transfer (HTTP/REST) synchronous requests or asynchronous message queueing protocols. In addition, each service should allow independent development and deployment, with the advantage of enhancing throughput and availability.

This independent deployment, together with the use of message communication patterns, such as request/reply, notifications or publish/subscribe, will provide us with a loose coupling architecture. The Remote Procedure Invocation (RPI) pattern is an alternative for the communications, with REST being one of the technologies most often used for this purpose. REST is an architectural style for distributed hypermedia systems, where services provide resources identified by URLs and where communications between REST services and their clients take place using HTTP main operations, mainly GET, POST, PUT and DELETE [21]. The decision between the use of a messaging broker or REST, or a combination of both, will depend on the commitment of availability versus simplicity, according to the project requirements.

A microservice architecture managing IoT devices must deal with the communication between the client and the services offered, in addition to the interoperability between the services. With this aim, web technology offers a series of mechanisms enabling the construction of heterogeneous systems with decoupled elements, whose parts are distributed on different platforms, are structured at different levels of abstraction, or use different communication protocols [22].

Based on web technology, the WoT paradigm [7,8,23] provides an abstract layer to make the communications and descriptions uniform under web standards and technologies. This approach considers a *thing* as an abstraction of a physical or virtual entity that provides interactions by using RESTful APIs. In this sense, TDs are homogeneous and must enable compatibility and interoperability with any kind of IoT solution. Furthermore, a WoT architecture should provide discovery mechanisms to allow clients to search and use things, thus facilitating a flexible and dynamic use of services, in addition to supporting their reuse.

*2.2. Complex event processing*

CEP [14] is a technology that allows us to capture, analyze and correlate a large amount of real-time data with the aim of detecting situations of interest in a particular domain [24]. In order to detect such situations as quickly as possible, we need to previously define a set of event patterns that specify the conditions to be satisfied by the analyzed events. These events can be simple (not a composition of other events) or complex (produced as a result of detecting an event pattern). A CEP engine is the software that permits the incoming data streams to be analyzed in real time, according to the defined patterns. Each CEP engine provides an Event Processing Language (EPL) to implement the patterns to be deployed in the engine.

CEP has been widely used within SOAs 2.0 to facilitate real-time monitoring of the events flowing in the SOA 2.0. For this purpose, the use of an ESB was required to transform the received events to the required data types, as well as the channel which routed the simple events from the data sources to the CEP engine and the complex events detected in the CEP engine to their final destination, either within the SOA itself or to third parties outside the SOA. Alternatively, some CEP engines offer the functionality of programming the engine using flow-based programming by supporting an internal event bus, which facilitates the use of the engine without the need to use another software tool, such as the ESB, for event routing.

Of the various existing CEP engines, we can highlight Esper [25] for its maturity and performance, as well as for the wide coverage of its Esper EPL provided for the pattern definition [26]. Specifically, Esper provides Esper Dataflows [27], which give support to the flow-based programming, thus removing the need to use an ESB to integrate event sources and event sinks of a CEP application. According to the Esper documentation, the use of dataflows is more efficient than the integration of an ESB with the Esper engine. The use of the Esper Dataflow CEP engine provides a variety of built-in operators than can be combined to program a dataflow, but is limited to certain incoming data formats and communication protocols. The supported data formats and communication protocols include Java Maps and Advanced Message Queuing Protocol (AMQP) [28], respectively, which will be used to illustrate the implementations in this paper.

*2.3. Messaging in smart scenarios*

In smart cities and smart applications in general, where large amounts of events are received, we need a software component that facilitates the communications and the management of input data. In such a scope, it is useful to use a message broker. Message brokers allow the source and destination of the messages to be kept completely decoupled through the implementation of an asynchronous communication mechanism. They also permit the messages to be stored in the broker until they can be processed by the target component, as well as other management facilities. In fact, each broker offers a series of facilities for communication, either through standard message queues or through a publish/subscribe mechanism making use of topics; these are the two most widely used models. Among the advantages of using message queues, we can mention their capacity to implement a load balance algorithm. In the case of message topics, every published message can be processed by all the consumers subscribed to that topic.

Among the messaging brokers we can highlight, for example, RabbitMQ [29], a lightweight, scalable and open source broker that is widely used. RabbitMQ supports several protocols: AMQP 0.9.1, STOMP, MQTT, AMQP 1.0 and HTTP and websockets. In particular, RabbitMQ was originally developed to support AMPQ 0.9.1 [30], where messages are published to exchanges, and exchanges distribute messages to queues using rules. The broker then delivers the messages to the consumers subscribed to the queues, or the consumers obtain the messages from the queues on demand.

**3. The microservice architecture at a glance**

The WoT CEP-enhanced microservice architecture, proposed in this work for smart cities in general and smart ports in particular, is represented in Fig. 1. Any real-time data processing architecture in a smart city scenario will have the following three key component categories: (1) data sources, (2) the data processor for the detection of situations of interest, and (3) target consumers of the situations of interest detected. Different implementations from the three categories, represented in Fig. 1, are described as things for exposing their interaction affordances and remain completely decoupled and are integrated through a message broker, as explained below:

**Data sources:** In the scope of smart cities, data sources often produce heterogeneous data in different formats and structures, such as JSON [31], XML [32] or CSV [33], as shown in Fig. 1. In addition, these data are shared through different communication protocols, such as AMQP [28] or MQTT [34], also in Fig. 1. In order to be able to process them in both streaming and real time, it will be necessary to previously transform the source data into a known data format and send them through a particular communication protocol.

**Transforming microservices:** These novel *transforming*

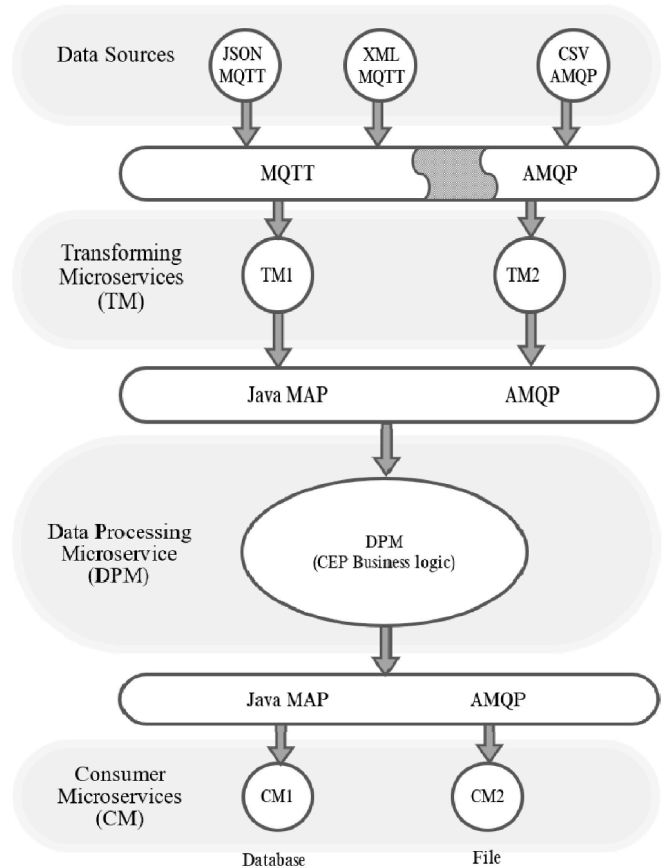

**Fig. 1.** Microservice-based architecture for smart scenarios.

*microservices* are proposed and implemented in this architecture to transform the incoming data with different data formats and protocols into the selected data type and protocol for the data processing microservice implementation: Java Maps [35] and AMQP [28] protocol, respectively.

**Data processing microservice:** This is the key microservice in the architecture. It uses CEP [14] to process the streams of incoming data to be able to notify the interested parties of the relevant situations detected in real time. This microservice permits the execution of the following operations at runtime:

- Providing the schema of the simple event to be received; such a schema can be automatically deployed in the CEP engine. Note that a schema defines the type of simple event together with its event properties.
- Providing the patterns for detecting the situations of interest; such patterns can be automatically deployed in the CEP engine.
- Providing the incoming source with the stream of incoming simple events and the output destination where the detected complex events must be submitted; such source and destination descriptions can also be deployed in the CEP engine.

**Consumer microservices:** These *consumer microservices* receive the output of the data-processing microservice and provide the interested parties with this output in the required format and protocol. Note that the outputs are the situations of interest detected as a result of the real-time processing of the incoming data.

As represented in Fig. 1, all these microservices are completely decoupled and communicate with each other through the asynchronous mechanism provided by a message broker. To facilitate such a communication with these and other microservices, all of them will provide their TD. As mentioned above, a TD exposes the interaction affordances of a thing in terms of properties, actions and events. Each operation of a microservice corresponds to an action and can be invoked using a POST method to the endpoint indicated by its TD. If an attribute used in a microservice as a local or a global variable can be relevant for the rest of the components in the architecture, its value can be offered as a readable, writable or observable property. A readable property is accessed by a GET method, whereas POST and PUT methods can be used for writing values in properties described as modifiable. The observer pattern for properties and the reception of events may be implemented with long polling, server-sent events or web sockets. With regard to event affordances, it must be noted that these kinds of interactions are intended to push data to other components and enable indirect communication by using an integration middleware, which, in our approach, is a message broker.

It should be noted that, although the architecture has been designed and implemented with a focus on IoT and smart city scenarios for IoT, it can be used in other scenarios not necessarily in the IoT domain, if required. However, our aims are (1) to reach real-time data processing of vast and heterogeneous amounts of data from different domain applications in the scope of smart IoT and WoT, (2) to maximize the reusability of the architecture components, and (3) to reduce resource consumption considering that many IoT scenarios require features such as low computation, limited memory and bandwidth that could otherwise be a bottleneck.

## 4. Microservices for the smart ports scenario

This section presents the microservices that we propose and implement for our microservice architecture to be used in smart port scenarios.

The following sections, describe firstly a smart port motivating scenario, and then, explain how these microservices should be connected to the external sources and destinations. The code of the implemented microservices is available at the Gitlab repository provided in the Supplementary Material section. Note that the code is generic and the microservices are, therefore, completely reusable in other domains apart from smart ports. Indeed, this is a key competitive advantage of the architecture presented in this paper.

### 4.1. Motivating scenario

Both road and maritime traffic currently produce a great deal of environmental pollution in seaports, affecting the cities in which they are located. Specifically, an ocean-going vessel emits an average of 18 tons of $CO_2$ during a port visit [36]. Overall, ships pollute 71% when they are berthed in port, 20% when they are maneuvering and 9% when they are at anchor [37].

One of the biggest problems to be dealt with is air emissions ($CO_2$, $NO_2$, $SO_2$, $PM_{2.5}$ and $PM_{10}$, among others), since pollution has a serious impact on health. For this reason, air quality plays an important role when it comes to the health of both port personnel and tourists, as well as citizens living near ports, which can lead to, or worsen, certain diseases or even cause death in certain risk groups (people with respiratory or cardiovascular diseases, among others) [38]. In this sense, controlling air quality is of great importance for people in general and for some groups in particular, such as the elderly, children, or people with heart conditions or allergies to certain substances that enter our body through inhalation.

In recent years, a number of international ports, such as Antwerp [36], Rotterdam [39], Hamburg [40], Amsterdam [41], and Singapore [42], and some Spanish national ports, such as Barcelona [37], Algeciras [43] and Palma [44], have evolved or are evolving towards smart ports. Although most of these ports monitor environmental quality, to the best of our knowledge, they do not focus on the automated sending of contextual alerts according to the situations of interest detected in real time.

Moreover, every port may receive air quality data from several sources; for instance, the port might receive (i) pollutant data from air quality stations placed in the vicinity of the port by the autonomous community administration, (ii) atmospheric data from central government weather stations, (iii) pollutant and atmospheric data from stations placed by the port authority itself within the port area. Thus, it would be useful to be able to choose between various reusable modules, depending on the types of data and input protocols, so that each port can easily configure its monitoring and notification system for air quality alerts in real time, which can lead to rapid actions to resolve the incidents.

Equally, each port may have different stakeholders interested in the results obtained from data processing. As examples, we can mention (i) the port authority itself, (ii) the operators who carry out the loading and unloading of materials in the port, (iii) the city council of the city where the port is located, and (iv) the citizens of that city. In addition, many of the stakeholders are likely to want to integrate these notifications into their existing software architectures, requiring various communication protocols and formats for sending the notifications. Therefore, the ideal scenario is to have a set of reusable modules from which to choose the one that fits our needs and that easily integrates with our detection system.

### 4.2. Source integration

To illustrate the microservice for the integration of data from different sources, we implemented a generic transformer that consumes messages from a specific AMQP queue and transforms JSON or XML data into Java Maps. This transformer microservice receives up to 5 arguments:

- *inputType* specifies the input type to consume. Supported types: JSON, XML.
- *inputQueue* specifies the input queue to consume the messages from.

- *outputQueue* specifies the output queue where the transformed messages are sent.
- *inputHost* specifies the source host from which the messages must be collected. Currently, it is expected to be an AMQP 0.9.1 broker.
- *outputHost* specifies the end host, where the transformed messages can be published. Currently, it is expected to be an AMQP 0.9.1 broker.

The first two arguments, *inputType* and *inputQueue*, are mandatory. The remaining arguments provide a default value: *outputQueue* is set to *map-events, inputHost* is set to *localhost* and *outputHost* is set to *localhost* by default.

The functionality of the data-transforming microservice is as follows:

1. **Connection configuration**: The microservice, by using the parameters previously explained, creates a *ConnectionFactory* object, which will establish the consumption of the information.
2. **Data consumption**: Once the connection is established, the microservice starts to consume messages from the *inputHost* and the *inputQueue*.
3. **Data transformation:** when a new message is consumed, it is automatically transformed from the *inputType* source to Java Map.
4. **Data publication**: once the data is homogenized to the input format required by the CEP engine, it is sent to the *outputHost* and published in the *outputQueue*.

Thanks to this process, we can transform several input data formats (JSON/XML) to Java Map, which is the input type required by the data processing microservice implemented. At this point, all the information, which has now been transformed and published in the output topic, is ready to be analyzed by the Esper CEP engine.

*4.3. Data processing*

The data-processing microservice provides an Esper CEP engine embedded in a Spring REST API. More specifically, this API offers three endpoints and internally manages all the interactions with the embedded CEP engine.

We make use of Esper dataflows [27] to process the data sources, through the immediate integration of AMQP 0.9.1 in the dataflow. Using dataflows, we can specify the host and the queue from which we wish to retrieve the information. This information is then published in an event bus that is internal to Esper, where the patterns, which have been previously deployed at runtime, are used to analyze and detect the situations of interest in real time.

The three endpoints provided by the REST API are as follows:

**POST /schema**

This receives an Esper schema in the JSON payload of the HTTP request and deploys it in the CEP engine, as shown in Listing 1.

*Listing 1 – EPL Schema definition*

```
{"schema": "@public @buseventtype create schema Dummy (p1 String, p2 Double)"}
```

Note that these schemas are mandatory to register the event type that we would subsequently like to analyze in our CEP engine.

**POST /pattern**

This receives an Esper EPL pattern in the JSON payload and deploys it in the CEP engine. Additionally, the action parameters are specified using @Tag(), as shown in Listing 2 (further details of these actions will subsequently be provided).

*Listing 2 – EPL pattern definition*

```
{"pattern": "@Tag(name=\"action\", value=\"file\") @Tag(name=\"name\", value=\"alert.txt\") @public @buseventtype select * from Dummy"}
```

The syntax in Listing 2 specifies that the pattern must be saved into a file named *alert.txt*. The required tags are: (i) *action,* which specifies the action, in this case the action is to save in a *file*; and (ii) *name,* which specifies the file name, in this case *alert.txt*.

*Listing 3 – EPL pattern definition to save alerts in a database*

```
{"pattern": "@Tag(name=\"action\", value=\"database\") @Tag(name=\"mongoURI\", value=\"mongodb+srv://<user>:<password>@<host>/test?retryWrites=true&w=majority\") @Tag(name=\"databaseName\", value=\"alerts\") @public @buseventtype select * from Dummy"}
```

Listing 3 shows another example of POST /pattern saving the alerts in a database. The syntax in Listing 3 specifies that the pattern must be saved into a NoSQL database. The required tags are: (i) *action*, which specifies the action, in this case saving in a *database*; (ii) *mongoURI*, which specifies the MongoDB URI to connect with —credentials included, in this case, *mongodb+srv://<user>:<password>@<host>/test?retryWrites=true&w=majority*; and *databaseName,* which specifies the database where the alerts will be saved, in this case *alerts*.

Finally, using the endpoint in Listing 4, we can specify and deploy any other kind of EPL definition, such as the context interval being used in the case study.

*Listing 4 – EPL pattern definition*

```
{"pattern": "@public create context IntervalSpanningSeconds start @now end after 300 sec;"}
```

All these patterns will be deployed in Esper public event bus. Thus when the data to be analyzed are received, these patterns are used to infer complex situations from such data. When a situation of interest is detected, it is sent, together with all the actions details, as previously explained, to an output queue. This queue is then used by the third microservice, as explained below.

**POST /dataflow**

This receives an EPL dataflow code in JSON payload and deploys it in the CEP engine, as shown in Listing 5.

*Listing 5 – EPL dataflow definition*

```
{"dataflow": "create dataflow AMQPIncomingDataFlow AMQPSource -> outstream<Dummy> {host: 'localhost', queueName: 'input-spring', collector: {class: 'AMQPSerializer'}, logMessages: true, declareAutoDelete: false} EventBusSink(outstream){}", "name": "AMQPIncomingDataFlow"}
```

The dataflow specifies the host, the queue and other parameters used to configure the consumer internally. This dataflow sets a consumption channel where the messages, which are already in Java Map format, are consumed and automatically published in the Esper event bus. The patterns are deployed in this bus, and are ready to be detected.

In the dataflow, we must specify the AMQP parameters required to consume the messages from a specific queue in a certain host. Other parameters can also be defined, such as the serializer, the queue durable property, etc. Esper dataflows allow developers to focus on the implementation of the patterns, relieving them of the struggles of implementing a way to send the information from the sources to the CEP engine. On occasions, this can be highly frustrating due to the limitations of the sources themselves.

Once the dataflow code is sent through the HTTP POST request, it is deployed in the CEP engine and the dataflow is instantiated and started in the CEP engine. The engine can then start to automatically consume the messages received as Java Map and analyze such information against the patterns previously deployed within it.

### 4.4. Output destination

Finally, once the situations of interest have been detected by the CEP engine and published to the output queue, it is time for the data consumers to act. To illustrate two different protocols for generating situation of interest notifications, we implemented the following microservice. It receives up to 2 arguments:

- *inputQueue* specifies the input queue to consume the messages from.
- *host* specifies the host where the AMQP 0.9.1 broker is running.

The first argument, *inputQueue*, is mandatory. The *host* is localhost by default.

When an alert is consumed, it checks which action has to be performed. This microservice provides two implemented actions:

- *File*: The detected alert is saved into a text file.
- *NoSQL Database*: The detected alert is saved into a MongoDB database.

These actions are specified when you send the pattern to the data processing microservice, as previously explained. Note that more actions, such as sending an email or publishing a tweet, could be coded within the microservice.

## 5. Case study and evaluation

This section discusses how the smart port case study was implemented through the composition of the microservices described in Section 4.

### 5.1. Configuration and deployment

This subsection shows how our three proposed microservices are composed to build the architecture for smart ports, where each service performs its work.

Bearing in mind the importance of controlling the levels of suspended particles $PM_{2.5}$ and $PM_{10}$ in ports, these are the values to be monitored with CEP in our architecture.

In order to configure and deploy the previously described microservices for this scenario, we should perform the following procedure. Our *SmartPort Transformer* microservice needs no configuration; it must simply be deployed, indicating, during the deployment, the input and output queues and host as well as the input type. It consumes JSON events, transforms them into Java Map and publishes them in the correct queue so that they are available to be consumed by the second microservice.

The second microservice, *SmartPort CEP*, then analyzes the information and detects the situations of interest in real time. As explained in previous sections, the CEP engine is embedded in a REST API, so the required configurations (event types, patterns and the dataflow) have to be provided by using HTTP POST requests. Listing 6 shows some of the EPL sentences provided to configure the CEP engine.

*Listing 6 – EPL sentences to case study evaluation*

```
(1) @public @buseventtype create schema AirQualityMeasurement as
    (PM10 integer, PM25 Double, stationId integer)
(2) create dataflow AMQPIncomingDataFlow AMQPSource ->
    outstream<AirQualityMeasurement>
(3) {host:'localhost',
(4) queueName: 'input-map',
(5) collector: {class:'AMQPSerializer'},
(6) logMessages: true,
(7) declareAutoDelete: false,
(8) declareDurable: true}
(9) EventBusSink(outstream){}
```

The first sentence in Listing 6 registers the *AirQualityMeasurement* event type, which has three event properties, one for the $PM_{10}$, another for the $PM_{2.5}$ and the station identifier. In the second sentence, the dataflow that consumes the events is created. Once these two EPL sentences are deployed, the CEP engine can recognize and consume *AirQualityMeasurement* events from the AMQP broker but, since there are no patterns deployed yet, no new alerts would be detected.

Thus, the following step consists of deploying the EPL patterns that allow us to have, at any time, the average value of the pollutants $PM_{10}$ and $PM_{2.5}$ for the last hour and for the last 24 h grouped by station. Another 4 patterns are then deployed for each pollutant in order to calculate their different air quality levels —Good, Moderate, Poor, Very Poor. Moreover, these level patterns have the *file* action configured, meaning that each time one of them is triggered, a text file with the complex event information detected is created. An excerpt of these patterns is shown in Listing 7; the remaining patterns of the case study can be found at the Gitlab repository specified in the Supplementary Material section. In particular, in Listing 7, we find (1) the code that fixes a context of 3600 s to more efficiently maintain the events in this period of time in memory; (2) the code that computes the hourly average $PM_{10}$ value for every station from the events received during the last 3600 s; (3) the code that computes the 24 h average $PM_{10}$ value for every station from the complex events created with the hourly averages, and finally (4) the code that sends an alert to a file every time a 24 h $PM_{10}$ average is in the range of a *good* value.

*Listing 7 – Sample of EPL statements to evaluate the air quality level for PM10*

```
(1) @public create context IntervalSpanningSeconds start @now end
    after3600 s;
(2) @Name('PM10_Avg1h_batch') @public context
    IntervalSpanningSeconds insert into PM10_Avg1h_batch select a1.
    stationId as stationId, avg(a1.PM10) as Value, count(*) as Total
    from AirQualityMeasurement a1 group by a1.stationId output
    snapshot when terminated;
(3) @Name('PM10_Avg24h_slide') @public @buseventtype insert into
    PM10_Avg24h_slide select a1.stationId as stationId, avg(a1.
    Value) as Value, a1.Total as Total from PM10_Avg1h_batch#time(24
    hour) a1 group by a1.stationId;
(4) @Tag(name='action', value='file') @Tag(name='name',
    value='alert.txt') @Name('PM10_Good') @public @buseventtype
    insert into PollutantLevel select'PM10'as kndAlrtDscr, 1 as
    AlertLevel, a1.stationId as stationId, a1.Value as Value from
    pattern [every a1 = PM10_Avg24h_slide (a1.Value >= 0 and a1.Value
    < 25)];
```

The last required configuration is the deployment of the third microservice, *SmartPort Actions*. We must indicate the input queue and host during the deployment. As explained, this microservice is in charge of receiving the alerts from the second microservice and performing the required actions. No additional configurations are required.

### 5.2. Thing description

The microservices are managed as configurable components and TDs are used to provide their interaction affordances to the rest of components in the architecture. Therefore, endpoints of the operations are exposed in JSON format to be automatically processable. As an example, Listing 8 shows the TD related to the *SmartPort Transformer* named as *EventTransformer*.

*Listing 8 – Thing Description of EventTransformer*

```
{
"@context": "https://www.w3.org/2019/wot/td/v1",
"id": "urn:dev:smartports:EventTransformer",
"title": "EventTransformer",
"securityDefinitions": {
"basic_sc": {"scheme": "basic", "in":"header"}
```

(*continued on next page*)

(*continued*)

```
},
"security": ["basic_sc"],
"actions": {
(1) "transformMessage": {
"title": "Transform message",
"description": "Transforms an input message into the desired output format",
"input": {
"type": "object",
"properties": {
(2) "event": {
"type": "string",
"description": "The message to be transformed"
},
(3) "json": {
"type": "boolean",
"description": "Indicates if the message comes as JSON"
},
(4) "xml":{
"type": "boolean",
"description": "Indicates if the message comes as XML"
}
}
},
(5) "output": {
"type": "map",
"description": "The transformed message as a Java Map"
},
(6) "forms": [{"href": "https://eventtransformer.smartports.ucase.es/transformm
    essage"}]
},
(7) "sendEventMap": {
"title": "Send event map",
"description": "Send the transformed message to the desired output topic",
"input": {
"type": "object",
"properties": {
"outputHost": {
"type": "string",
"description": "The output host where the Message Broker is running"
},
"outputQueue": {
"type": "string",
"description": "The output topic where the message will be sent"
},
"eventMap": {
"type": "map",
"description": "The message, as Java Map, to be sent"
}
}
},
(6) "forms": [{"href": "https://eventtransformer.smartports.ucase.es/sendeventmap"}]
}
}
}
```

The *EventTransformer* microservice provides two actions in its TD, as shown in Listing 8: the *transformMessage* operation in code block (1) to transform an input message into the desired output format, and the *sendEventMap* operation in code block (7) to send this transformed message to topic introduced as a parameter. The TD defines the endpoints — see code block (6)— to enable the automatic discovering and invocation of such operations. Furthermore, each action is described with the method signature. For example, the *transformMessage* receives three parameters as an input: the message to be transformed —code block (2), an optional parameter if the message has a JSON format —code block (3), and an optional parameter if the message has an XML format — code clock (4). This operation generates the transformed message in a Java map as an output —code block (5). As the descriptions of each component are in TD form, it is possible to implement automatic processes that carry out communication between things. In addition, this facilitates the reuse and replacement of components, since decoupled elements are being accessed through the APIs that they expose.

### 5.3. Simulation

With the aim of testing the functionality of the microservice composition, we simulated real behavior for two air quality stations providing data for $PM_{2.5}$ and $PM_{10}$. We could have simulated an unlimited number of stations, but there are not likely to be a large number of these in the area occupied by a seaport. For example, the port of Cadiz, located in the city and the bay of Cadiz, could monitor air quality at one station in the city of Cadiz, and another station in Puerto Real, a town in the bay where port activities are also carried out.

For this simulation, we used nITROGEN [45], which permits the generation of synthetic data for the IoT. In nITROGEN, there are several connectors that allow us to send the generated data to a Kafka topic, an AMQP broker, a MQTT broker, or a file, among others. In this case, we used the AMQP 0.9.1 connector to send the simulated events to the RabbitMQ server.

In particular, the simulator was configured with a simulation composed of two stations, one for the Cadiz location and the other for the Puerto Real location. Both stations sent simulated $PM_{2.5}$ and $PM_{10}$ data to a RabbitMQ queue.

When the simulation was run, the following process was conducted, as depicted in Fig. 2:

1. nITROGEN started producing JSON events and submitted them to the RabbitMQ queue.
2. *SmartPort Transformers* consumed these JSON events from the queue, transforming them into Java Map events and publishing them in the specified output queue, where *SmartPort CEP* consumed the data.
3. *SmartPort CEP* consumed the Map events (the JSON events homogenized) from the previously mentioned queue and analyzed them against the patterns previously deployed.
4. The complex events detected by the CEP engine were published in the alerts output queue.
5. *SmartPort Actions* consumed the alerts detected from the alerts output topic and performed the required actions.

### 5.4. Performance evaluation

Next, we evaluated the performance of the proposal to show its suitability for IoT scenarios in which it is required to consume, process and analyze large amounts of data in real time.

For this purpose, we tested two of the three microservices previously

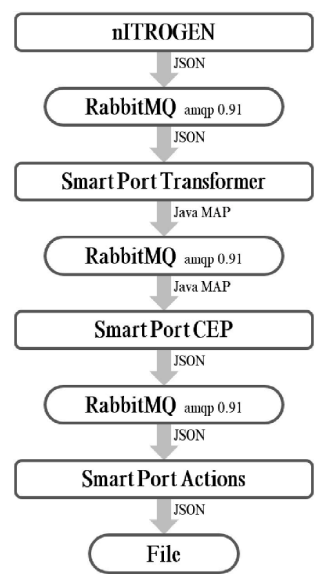

**Fig. 2.** Data flow and microservices in the Smart Port scenario.

proposed:

- *SmartPort Transformers*: By testing it, we could check that the microservice was able to consume, process and transform large inputs of information in real time.
- *SmartPort CEP*: By testing it, we could ensure that the CEP engine was able to handle these large amounts of data in real time.

The performance of the *SmartPort Actions* microservice did not require evaluation, as it will only be used when complex events are detected, and we do not expect to have a large number of complex events detected per second.

In the following lines, two performance tests carried out for *SmartPort Transformers* and *SmartPort CEP* are presented, showing their capacity to deal with high load scenarios. The first set of tests, as later explained, was carried out for a *short* period of time (10 min) with the aim of finding the maximum incoming rate accepted by the system. Subsequently, a *longer* test lasting 12 h with the maximum incoming rate supported by the system was performed to verify that such a load could be supported for longer periods of time

The computer used to test these microservices had the following specifications: Windows 10 Enterprise, 64 bits, AMD Ryzen 5 3600X,

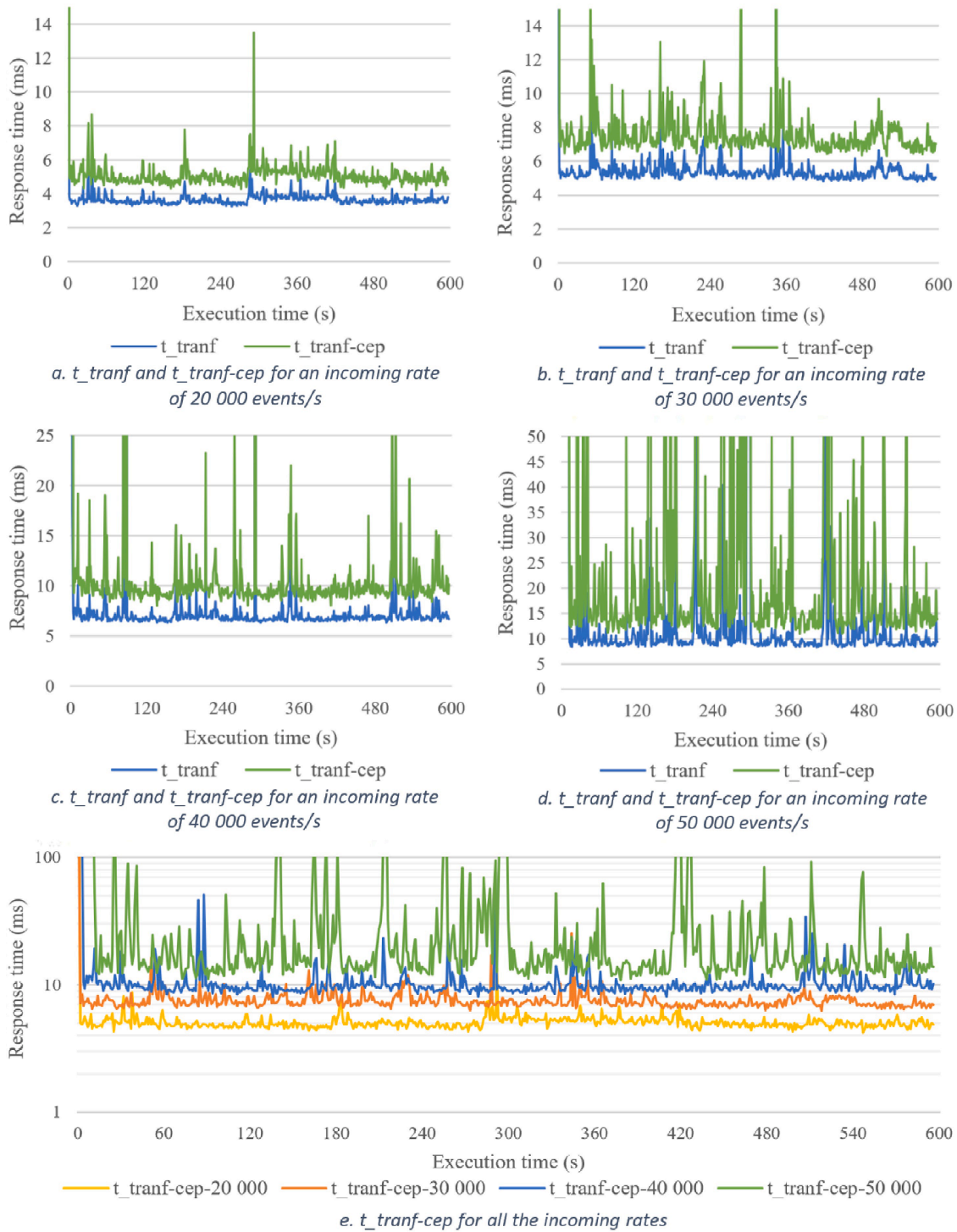

**Fig. 3.** Response and communication times for transformation and CEP microservices for increasing incoming rates.

16GB RAM, and 1 TB HDD. In this computer, the microservices were deployed along with an AMQP 0.9.1 broker, which in this case was RabbitMQ 3.8.2.

For the testing, we adapted the case study patterns and deployed them in the SmartPortCEP microservice. In particular, to limit the time cost of running the tests, the average values were now calculated every 5 and 10 min seconds for the *short* tests, so we would have the air quality levels —Good, Moderate, Poor, Very Poor— for the last 600 s. The full definition of the EPL schema, dataflow and patterns used in these tests can be found in the Mendeley dataset referenced in the Supplementary Material section. After deploying the pattern, the JSON events with random values of $PM_{2.5}$ and $PM_{10}$ were submitted to the AMQP queue. We tested both microservices separately and then together. We successively increased the ratio of input messages to the system until the limit was reached. The input message arrived in JSON format at the *SmartPort Transformers* microservice, where it was transformed into Java Map.

We measured the response time in this microservice, being $t_{tranf}$ the time passed from the generation of the JSON event in the simulator until the transformation in the microservice was carried out; let us remember that after its generation in the simulator it was sent to the queue in RabbitMQ to which the microservice had previously subscribed. Moreover, we also measured the response time when connecting the execution of the two microservices —*SmartPort Transformers* and *SmartPort CEP*. Therefore, being $t_{tranf-cep}$ the time passed between the generation of the JSON event in the simulator and a complex event being detected in the *SmartPort CEP*, which includes the communication time between the simulator and the message queue to which *SmartPort Transformers* is subscribed, the processing time in *SmartPort Transformers*, and the communication time from *SmartPort Transformers* to the message queue to which *SmartPort CEP* is subscribed and from the named queue to the latter, as well as the processing time in *SmartPort CEP* (see Fig. 2).

Note that having such high ratios of messages per second, the graphical representation of all $t_{tranf}$ and $t_{tranf-cep}$ would be too dense. Thus, for purposes of clarity, we calculated the mean value of the $t_{tranf}$ and $t_{tranf-cep}$ values for all events sent in each second, and these are the values shown in Fig. 3. Specifically, Fig. 3(a), (b), (c) and (d) show the mean values of $t_{tranf}$ and $t_{tranf-cep}$ for an incoming rate of 20 000, 30 000, 40 000 and 50 000 events per second, respectively. As can be seen, the time elapsed between the event being generated in the simulation and it being transformed to Java Map in the first microservice —$t_{tranf}$— is kept below 5, 7, 10 and 30 ms, respectively for each input ratio, except for the warm-up and the occasional moment of processor overload. Furthermore, if we add to that time the time it would take to receive the transformed event in the second microservice and have it processed by the CEP engine —$t_{tranf-cep}$, we see they are kept below 8, 12, 25 and 50 ms, respectively for each input ratio, except for the warm-up and the occasional moment of processor overload. Such performance rates are reasonably good for the high input ratios we are testing, and more than sufficient for our case study. Finally, Fig. 3(e) compares $t_{tranf-cep}$ for all the tested incoming rates in logarithmic scale; we can observe that the response times are quite stable for 20 000, 30 000 and 40 000 input rates. However, in the case of 50 000 events per second, although supported by the system, many alterations in the response time are already noticeable.

Additionally, Fig. 4 better illustrates the percentages of data with low response time rates. As we can see, for an input rate of 20 000 events per second, most of the data are processed in less than 2 ms (data in light blue). However, this percentage successively decreases when input rates are increased to 30 000, 40 000 and 50 000 events per second. Despite the increase in response times, we can see that, for 30 000 and 40 000 events per second, most of the data is maintained with more than adequate response times, and it is at 50,000 events per second where we start to see peaks of relatively high rates. In this case, a high percentage of data remains within between 10 and 40 ms of response time along with some additional peaks over this rate.

As previously mentioned, as the amount of data generated is so high and we measured the response time for each of the data, the results cannot be correctly appreciated in a single graph. For this reason, for the representation in Fig. 3, we calculated the mean response time for all the events in each second of execution. The full set of data obtained in the performance evaluation can be found in the Mendeley dataset referenced in the Supplementary Material section. In addition, in Fig. 5(a), (b), (c), and (d), we can see the minimum (Min.), maximum (Max.), mean (Mean) and the standard deviation values subtracted and added to the mean (Mean-SD, Mean+SD) for all the events in every second of execution in a logarithmic scale. The standard deviation shows that the performance times dispersion is reasonable and not too far from the mean.

Besides, we also measured the CPU and memory consumption during the tests. As shown in Fig. 6 the percentage of CPU usage at an incoming rate of 20 000 events per second is below 50%; however, it increases to almost 80% at 30 000 events per second and around 85 and 90% for 40 000 and 50 000 events per second, respectively. In these last 2 cases it is evident that the CPU consumption is already excessive and the system is at the limit of its capacity. However, Fig. 7 shows how the memory consumption, although slightly increasing as the input message rate increases, stabilizes within reasonable limits: around 7.5 GB consumption for input rates of 20 000 and 30 000 events per second and 8 GB for 40 000 and 50 000 events per second.

We also conducted a long-term test with an input ratio of 30 000 events per second, in which the system responded properly. A total of 1 283 995 749 events were processed by the system in 12 h, the average response time was 0.03 s, and the average consumption of CPU and RAM memory were 49.5% and 6.23 GB, respectively. The EPL schema and patterns used for the *long* test were extended. In this way, the average values are now calculated every hour and 2 h, with the aim of collecting and analyzing this enormous amount of data. The full definition of the EPL schema, dataflow and patterns used in this test together with the full log file of the test are available at the Mendeley repository referenced in the Supplementary Material section.

Finally, we have tested how the system would behave on a less powerful machine and we have deployed the microservices on a Raspberry PI Broadcom BCM2711, Quad core Cortex-A72 (ARM v8) 64-bit SoC @ 1.5 GHz with 4GB RAM. In this case we have made the tests for the following incoming rates: 5 000, 7 5000, 9 000 and 10 000 events per second. Fig. 8 shows the response times for the increasing events incoming rates. As shown in the figure, the Raspberry can properly deal with incoming rates of 5 000 and 7 500 events per second, but the response times are unacceptable for an incoming rate of 9 000 events per second and the system can definitely not deal with 10 000 events per second of incoming rate. Additionally, Fig. 9 better illustrates the percentages of data with low response time rates. As we can see, for an input rate of 5 000 events per second (see Fig. 9(a)), most of the data are processed in less than 2 ms (data in light blue and dark orange). However, this percentage decreases when input rate increases to 7 500 events per second (see Fig. 9(b)), but still around 60% of the data is processed in less than 2 ms. However, when we increase the rate to 9 000 events per second, the average response time is always over 160 ms (all the chart would be dark blue) and the system stops working in the middle of the test for an incoming rate of 10 000 events per second.

## 6. Related work

Over recent years, several works [46–49] have emphasized the importance of integrating IoT with smart ports to easily share data between port agents.

For the seaport of Las Palmas de Gran Canaria, Spain, Fernández et al. [50] proposed a platform for supporting sensor data monitoring. It uses the FIWARE ecosystem to store and notify the alerts related to the management of containers and the state of the sea. For the seaport of Valencia, Spain, Sarabia-Jácome et al. [51] implemented a seaport data space also based on the FIWARE ecosystem to share data and facilitate

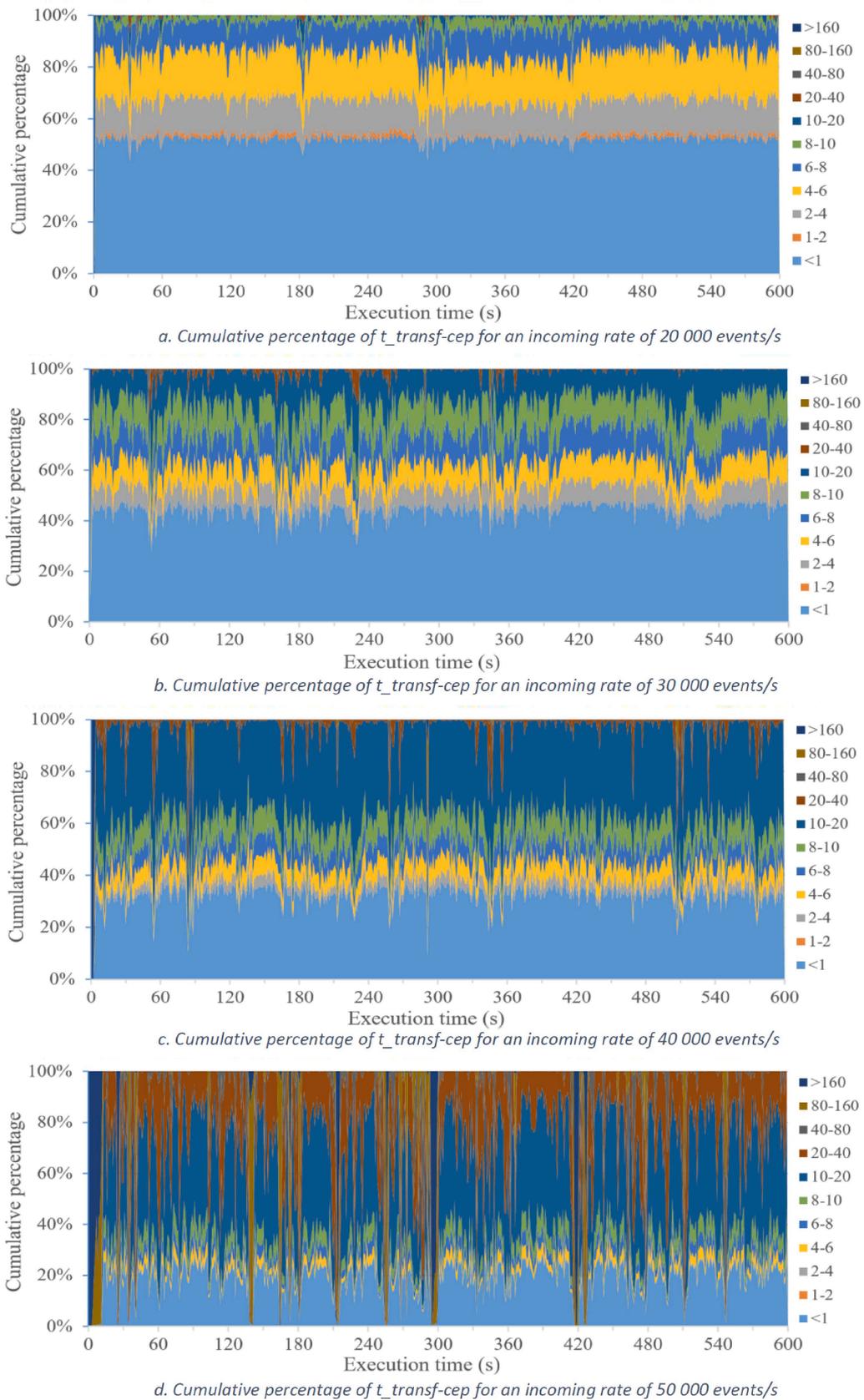

**Fig. 4.** Cumulative percentage for response and communication times for the sequential execution of transformation and CEP microservices for increasing incoming rates.

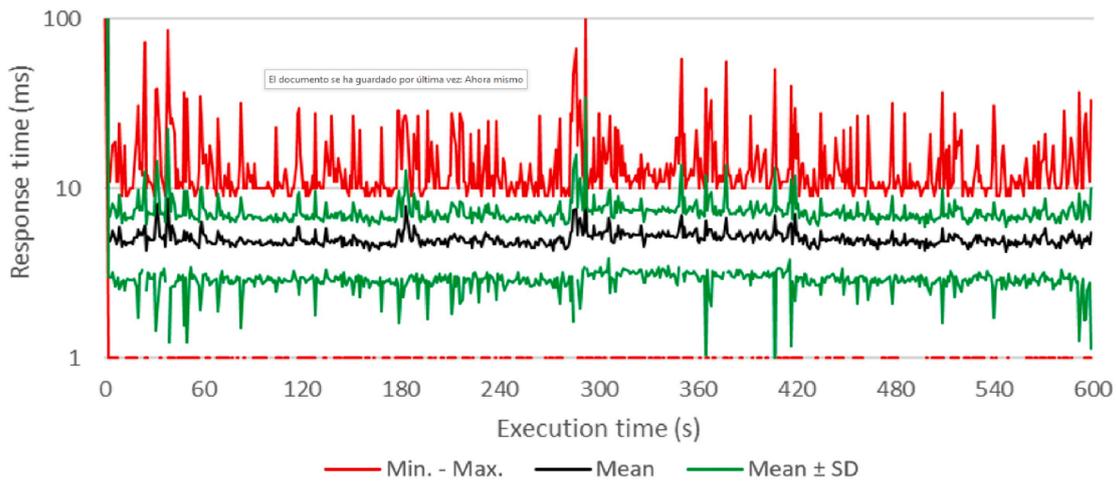

*a. Minimum, maximum, mean and standard deviation of t_tranf-cep for an incoming rate of 20 000 events/s*

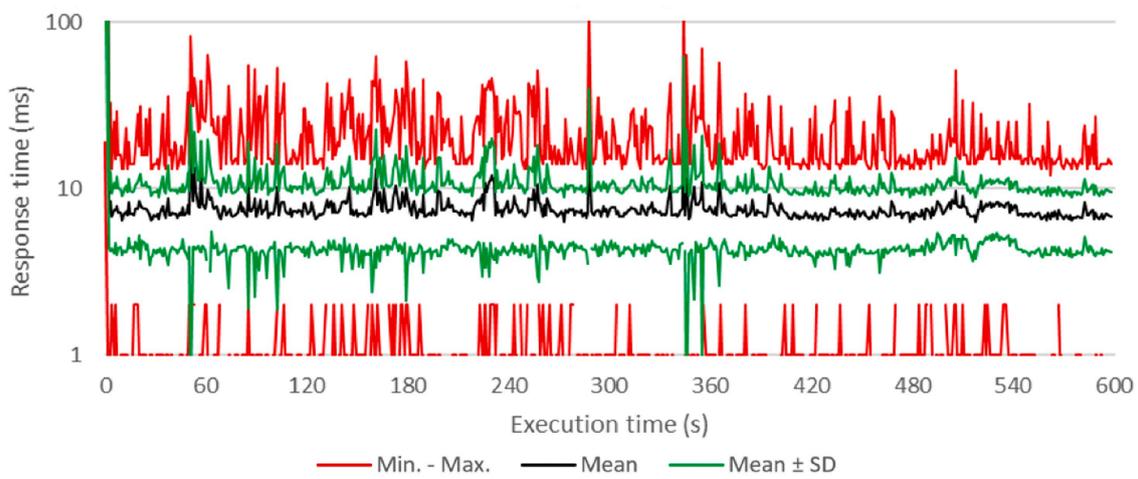

*b. Minimum, maximum, mean and standard deviation of t_tranf-cep for an incoming rate of 30 000 events/s*

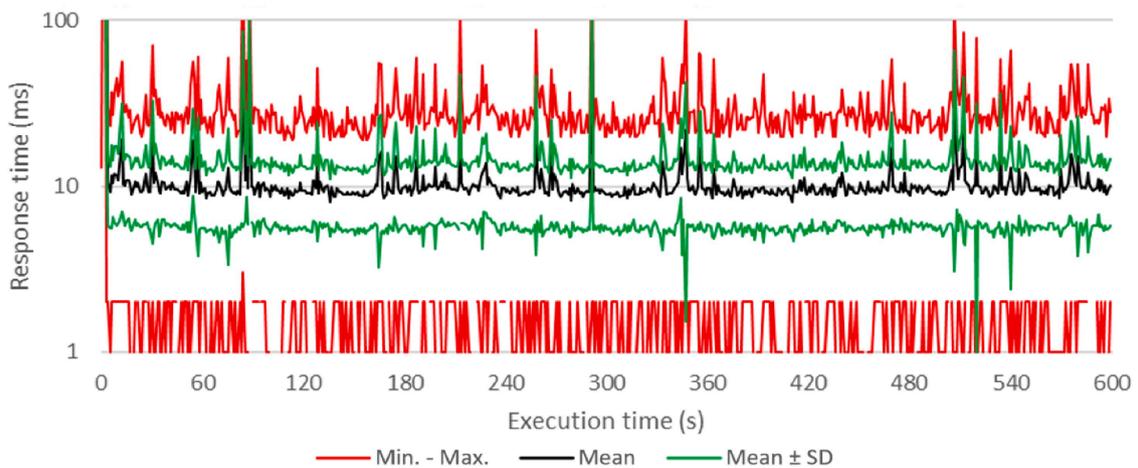

*c. Minimum, maximum, mean and standard deviation of t_tranf-cep for an incoming rate of 40 000 events/s*

**Fig. 5.** Minimum, maximum, mean and standard deviation for response and communication times for the sequential execution of transformation and CEP microservices for increasing incoming rates.

the interaction between stakeholders. This architecture is integrated with a large data platform implemented with Apache Spark to analyze data on terminal operations and vessel positions. For the seaport in Le Havre, France, Rajabi et al. [52] proposed a smart port architecture composed of four layers: port infrastructure & stakeholders, enabling technologies, smart port services and smart port goals. They implemented a Java application to decode Automatic Identification System (AIS) messages, which are the main data sources of the application.

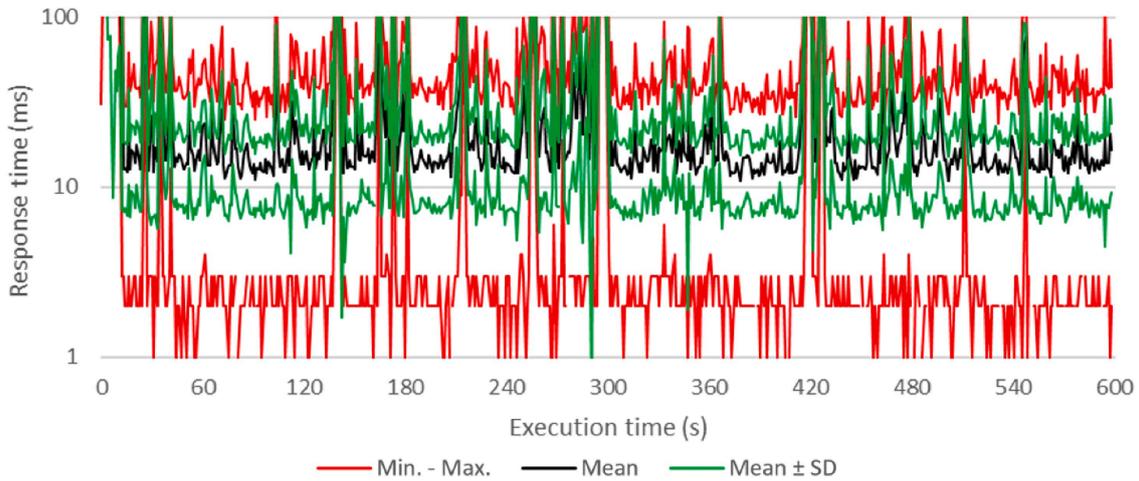

d. Minimum, maximum, mean and standard deviation of t_tranf-cep for an incoming rate of 50 000 events/s

**Fig. 5.** (*continued*).

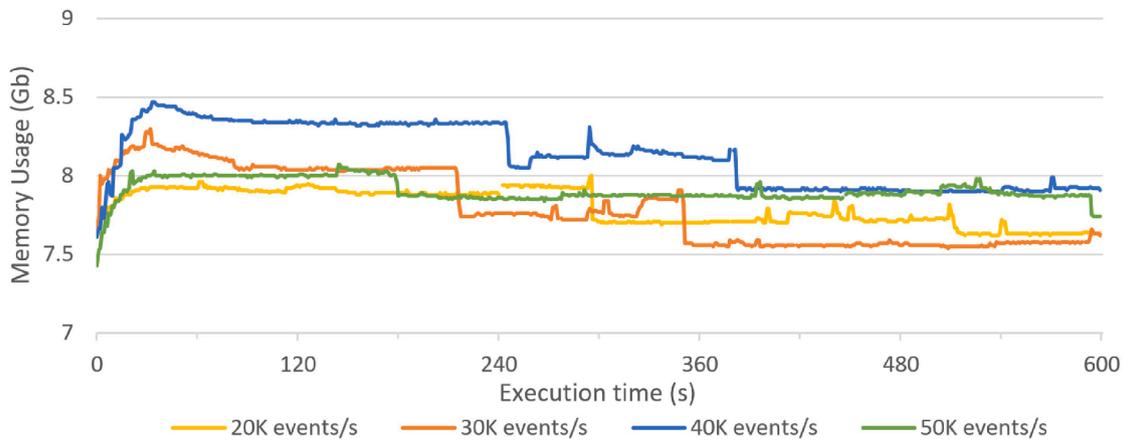

**Fig. 7.** Memory consumption for the sequential execution of transformation and CEP microservices for increasing incoming rates.

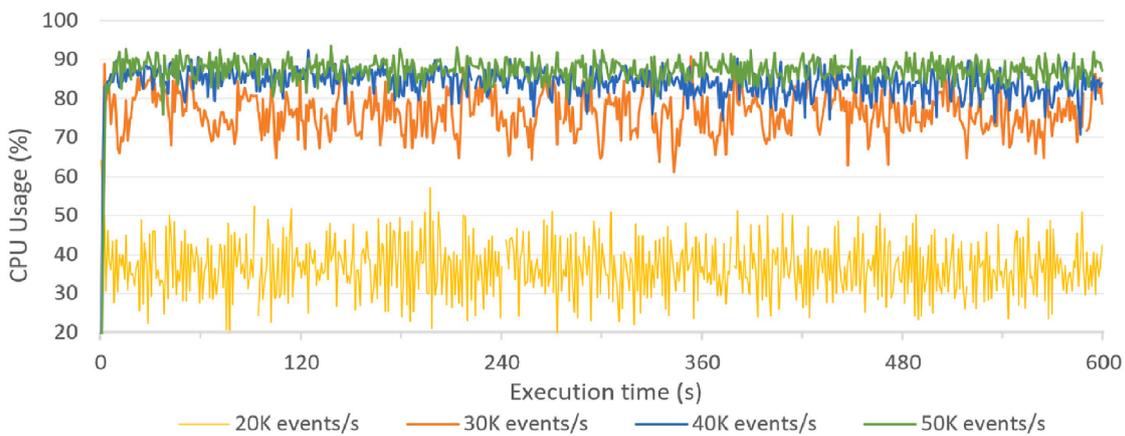

**Fig. 6.** Percentage of CPU consumption for the sequential execution of transformation and CEP microservices for increasing incoming rates.

These decoded messages are then stored in a MongoDB database. This stored data can be analyzed through a web-based interface implemented in Ruby on Rails. In addition, Ferretti and Schiavone [53] explained how IoT is essential to redesign and improve the efficiency and effectiveness of business processes in the seaport in Hamburg, Germany. For seaports in Indonesia, Kusuma and Tseng [54] designed an IoT platform that can be accessed by port services users with the aim of assisting the service process. However, it is, as yet, unimplemented and no information is provided about which technologies should be used to process real-time data and notify alerts to the interested stakeholders.

It is worth noting that none of the above works benefits from a WoT microservice architecture integrated with CEP technology, as we propose in the present work. This system not only makes data sharable and reusable between ports in a standardized way around the world, but is also capable of processing and correlating heterogeneous real-time data as well as automatically detecting air quality event patterns to be notified to the interested port agents and citizens. Additional situations of interest in the smart ports could also be notified, deploying new event types and patterns to the microservices. Furthermore, such microservices can be available in WoT standardized repositories to facilitate their discovery and reuse.

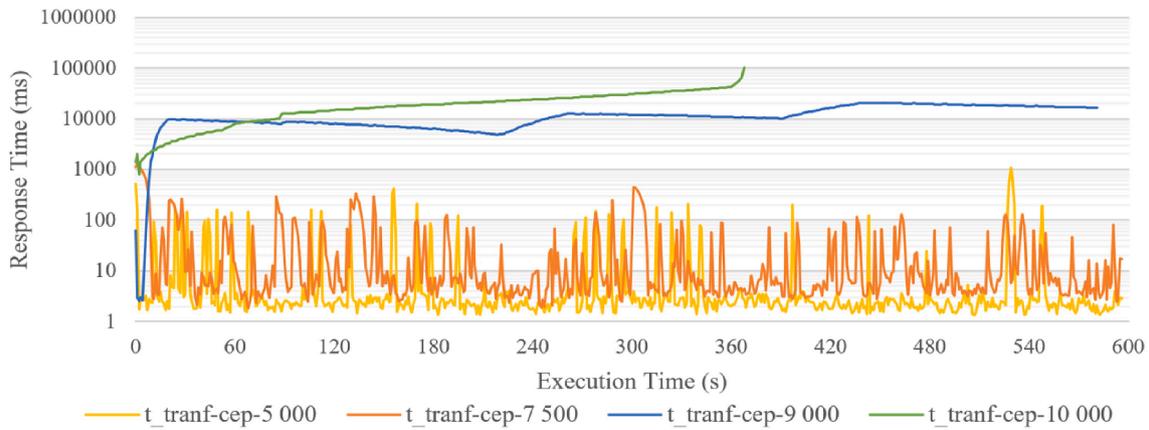

**Fig. 8.** Response and communication times for transformation and CEP microservices for increasing incoming rates in a Raspberry PI.

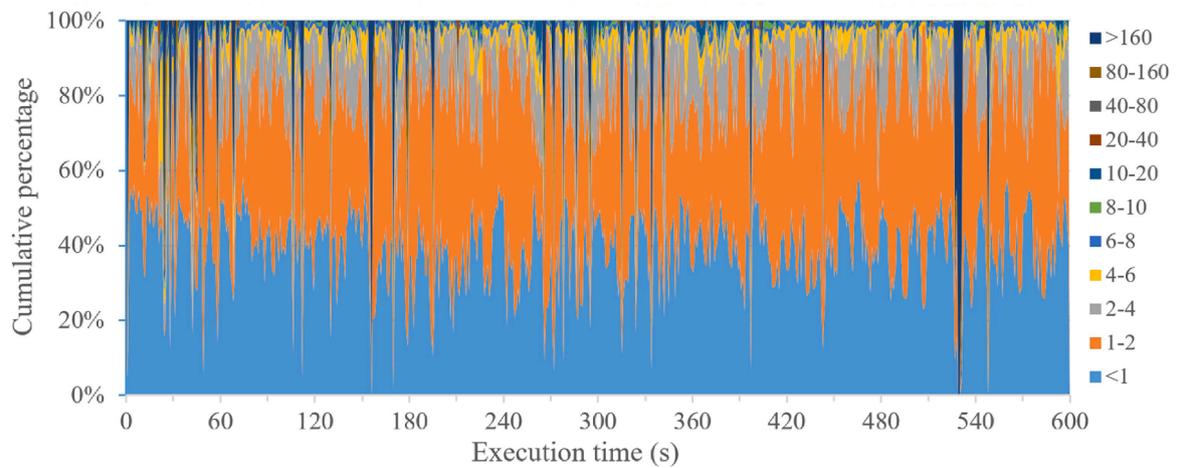

**Fig. 9.** Cumulative percentage for response and communication times for the sequential execution of transformation and CEP microservices for increasing incoming rates in a Raspberry PI.

In other domains, such as smart cities, works have proposed the integration of microservices and CEP. As an example, Scattone and Braghetto [55] describe a microservice architecture for distributed CEP in smart cities. In particular, they propose the use of Esper as CEP engine, Ruby on Rails for the microservice and RabbitMQ for transmitting events in an asynchronous way. However, the architecture implementation is still an ongoing work.

Chegini and Mahanti [56] envisioned a microservice framework for the development of IoT-based context-aware intelligent decision-making systems. Its main challenges are real-time big data processing, managing the heterogeneity of data, software and hardware, and increasing fog resiliency. This framework, which is proposed, but not implemented, suggests CEP as a candidate technology to adapt agents in the microservice architecture to changing conditions.

Finally, as mentioned in Section 1, some of the authors of this paper [17] proposed a microservice architecture integrating CEP and machine learning technologies to provide real-time context-aware actions based on predictive data streaming processing. While the work in [17] makes use of an ESB integrated with the Esper CEP engine, the present work uses no ESBs, but implements the functionalities provided by an ESB, such as data transformation and data routing, through microservices. Therefore, the microservice architecture proposed in the present paper is lighter than the one in [17] and more appropriate to be deployed to interact with other microservices.

## 7. Discussion

If we go back to the challenges and needs discussed in Section 1, we will see how our proposed solution meets the expectations of all of these, as explained below.

Regarding the ability to process *heterogeneous data*, the advantage of using a microservice architecture lies in the fact that the functionality can be quickly extended by deploying and interconnecting a new service. In this sense, our case study shows a microservice that transforms incoming JSON data into Java Map format. In addition, we can specify at runtime the structure of the new input data the microservice will receive and we can implement and deploy other microservices. These microservices, having different data sources, make the transformations that such data requires for its homogenization prior to sending it to the processing microservice. The same applies to notifications or alerts issued by the system once situations of interest are detected after data processing: we can add new microservices for other types and/or recipients of notifications without the need to stop the system.

Secondly, the provision of a microservice that integrates a CEP engine provides a large real-time data processing capability. In addition, we can add new patterns to be detected at runtime, as well as new microservices to which to send notifications or alerts of events of interest detected by the engine. Given this feature of the system, and those described in the previous paragraph, we can affirm that the system is perfectly *scalable and maintainable* in terms of its functionality and evolution. Furthermore, thanks to the use of the WoT paradigm, which homogenizes access to these services and the interoperability of the system by creating an abstract layer based on web technologies through the microservice TDs, we can ensure that the *reusability* of the services provided is guaranteed. Furthermore, thanks to the TDs incorporated, the components will be easily discovered in WoT repositories and code for the interaction with other components can be automatically generated.

Moreover, the performance evaluation of the case study, using a synthetic data simulator to generate large data rates per second, has shown that the system responds with very reasonable response times for input rates of 20 000, 30 000 and 40 000 events per second, while maintaining reasonable memory consumption, although CPU consumption is high in the latter two cases. It is only at 50 000 events per second that we start to see the system taking slightly longer to respond, although still within acceptable limits depending on the needs of the particular case study. In this sense, the system *scales efficiently* in terms of the size of the incoming data stream and the time required to process it. When testing on a Raspberry Pi, we found that we obtained a reasonable system response up to an input rate of 7500 events per second.

We would like to emphasize that, although the solution proposed here aims to meet the demands in terms of low processing power, memory and limited bandwidth that can be found in some IoT scenarios, without reducing the interoperability of the solution or the real-time processing of large amounts of data that can occur in these scenarios, the architecture could be used in any other scenarios where the aim is to have a modular, reusable and easily maintainable architecture for monitoring data from heterogeneous sources in real time and sending notifications to different stakeholders.

Last, but not least, as seen in Section 7, to the best of our knowledge, none of the existing proposals for smart ports provides a fully interoperable approach, as we do with the use of a WoT microservice architecture, nor do they benefit from an integrated CEP engine to facilitate real-time detection and notification of situation of interest. Moreover, although we can find other approaches that benefit from the use of microservice architecture in the IoT and smart cities domain, most of them lack the advantage of interoperability, while also not having a fully standard microservice with an integrated CEP engine. It should be noted that our proposal can be used to detect and notify additional situations of interest in smart ports, deploying new event types and patterns to the microservices. It can also be reused for other domains, such as smart industry, smart health, smart energy, etc. [49,57,58].

## 8. Conclusion

This paper shows how a microservice architecture can transform and process large amounts of heterogeneous data coming from the IoT in a smart city environment, and, more specifically, in a smart port. Moreover, describing microservices by following the WoT standards will facilitate their reuse by other developers and their interoperability with other microservices of existing or new projects. It is undoubtedly a solution that keeps pace with the rapid evolution of technologies and telecommunications in general and the IoT in particular, and which solves major challenges for its application in different domains in the field of smart cities.

In our future work, we will extend the number and variety of *things* available in the proposed architecture. In this sense, we will implement new, alternative microservices (encapsulated and described as *things*) for transformers, data processors and consumers, in addition to the gateways and controllers related to the IoT devices. The set of *things* to be used in the architecture will be stored in a repository to be inspected and so discover the most suitable *things* depending on the context information or the application domain [59].

**Supplementary materials**

We have uploaded JAR files for the Smart Port microservices together with the EPL schema and patterns for the case study, the EPL schema and patterns used for the performance evaluation and the data collected from such evaluation at Mendeley Data. The dataset can be found at URL https://doi.org/10.17632/2g2s3ybcmn.1. The supplementary material is composed of the following folders and items:

Smart port microservices: This includes the jar files for the three microservices —*SmartPortTransformers, SmartPortCEP* and *SmartPortActions*, the instructions for their deployment —readme.md file— and the schema and patterns defined for the case study —SmartPortSchemaAndPatterns.txt.

Performance EPL schema and patterns: This includes the Esper EPL schema and patterns defined both for the short performance tests as well as for the long ones.

Performance evaluation results data files: This includes the spreadsheet response time values obtained from every performance test both for the short performance tests as well as for the long ones.

The code of the Smart Port microservices can be openly accessed through a Gitlab repository available in the following link: https://gitlab.com/ucase/public/SmartHealthyPort.

**CRediT authorship contribution statement**

**Guadalupe Ortiz:** Conceptualization, Methodology, Writing – original draft, Writing – review & editing, Funding acquisition. **Juan Boubeta-Puig:** Conceptualization, Methodology, Writing – original draft, Writing – review & editing, Funding acquisition. **Javier Criado:** Conceptualization, Methodology, Writing – original draft, Writing – review & editing, Funding acquisition. **David Corral-Plaza:** Software, Validation, Writing – review & editing. **Alfonso Garcia-de-Prado:** Validation, Visualization, Writing – review & editing. **Inmaculada Medina-Bulo:** Conceptualization, Writing – review & editing, Funding acquisition. **Luis Iribarne:** Conceptualization, Writing – review & editing, Funding acquisition.

**Declaration of Competing Interest**

The authors declare that they have no known competing financial interests or personal relationships that could have appeared to influence the work reported in this paper.

**Acknowledgments**

This work was supported by the Spanish Ministry of Science and Innovation and the European Regional Development Fund (ERDF) under research projects CoSmart and FAME (ref. TIN2017–83964-R and RTI2018–093608-B-C33), and by regional projects (ref. CEIJ-C01.1 and CEIJ-C01.2) coordinated from UAL-UCA and funded by Campus de Excelencia Internacional del Mar (CEIMAR) consortium. We would like to thank J. M. Pérez Sánchez, A. M. Garrido López and R. J. Catalán Alonso from the Autoridad Portuaria de la Bahía de Cádiz, Spain, for sharing valuable knowledge on port management and environment and collaborating with us in this work.

**Guadalupe Ortiz** obtained her Ph.D. in Computer Science at the University of Extremadura (Spain) in 2007, where she worked from 2001 as Assistant Professor. In 2009 she joined the University of Cadiz (UCA) as tenured Associate Professor in Computer Science and Engineering. Her research interests focus on the integration of complex-event processing and context-awareness in service-oriented architectures in the Internet of things.

**Juan Boubeta-Puig** received the Ph.D. degree in computer science and engineering from the University of Cadiz (UCA), Cádiz, Spain, in 2014. He is an Associate Professor with the Department of Computer Science and Engineering, UCA. His research interests include real-time big data analytics through complex event processing, event-driven service-oriented architecture, Internet of things, blockchain and model-driven development of advanced user interfaces, and their application to smart cities, industry 4.0, e-health, and cybersecurity. Dr. Boubeta-Puig was honored with the Extraordinary Ph.D. Award from UCA and the Best Ph.D. Thesis Award from the Spanish Society of Software Engineering and Software Development Technologies.

**Javier Criado** is an Assistant Professor at the Department of Informatics, University of Almería, (UAL), Spain. In 2009, he joined the Applied Computing Group (TIC-211), a research group in the UAL. He has participated in four national research projects (refs. TIN2007–61,497, TIN2010–15,588, TIN2013–41,576-R and TIN2017–83,964-R) and two regional research projects (refs. P10-TIC6114 and CEIJ-C01.2). From 2011–2015, he was supported by an FPU grant (Ref. AP2010–3259). He received his PhD (2015) in Computer Science from the UAL. His research interests include: UML design, Model-Driven Engineering, Component-Based Software Engineering, User Interfaces, Interoperability, Service-Oriented Architectures, Microservices and Web of Things.

**David Corral-Plaza**, received the Ph.D. degree in Computer Science from the University of Cadiz (UCA), Spain, in 2021. Since 2017, he has been a Predoctoral Researcher at the Department of Computer Science and Engineering, UCA, developing his Ph.D. and collaborating in lecturing at the Degree in Computer Science of the UCA. His research focuses on the integration of Complex Event Processing and Stream Processing, Event-Driven Service-Oriented Architectures, and microservices for heterogeneous data processing in the Internet of Things domains.

**Alfonso García-de-Prado** received his Ph.D. degree in Computer Science at the University of Cadiz (UCA), Spain, in 2017, where he has been an assistant professor since 2012. Previously he had been a developing programmer, analyst and consultant for international industry partners. His research focuses on context-aware service-oriented architectures, as well as their integration with complex event processing and the Internet of things.

**Inmaculada Medina-Bulo** received the Ph.D. degree in Computer Science from the University of Seville, Spain. She has been with the Department of Computer Science and Engineering, University of Cadiz, Spain, since 1995. She has published more than 150 peer-reviewed papers, participated in conference and workshops organization, and acted as a reviewer for several journals. She is the head of the UCASE Software Engineering Research Group, of the Spanish Research Network in Search-Based Software Engineering and of other Research Project in Software Engineering. Her current research interests include software testing, search-based software engineering, the IoT, CEP, and SOA 2.0.

**Luis Iribarne** is an Associate Professor at the Department of Informatics, University of Almeria (Spain), and Head of Applied Computing Group (ACG). He received his BSc and MSc degrees in Computer Science from the University of Granada, and his Ph.D. degree in Computer Science from the University of Almeria, and driven at the University of Málaga (Spain). From 1991 to 1993, he worked as a Lecturer at the University of Granada. Since 2006, he has served as the Principal Investigator (PI) for seven R&D projects founded by the Spanish Ministry of Science and Technology. He has also acted as supervisor for funding agencies in Spain (ANECA, ANEP and AEI) and Argentina. His main research interests include simulation, component-based software development, model-driven engineering, and software engineering.